%% file: Main.tex
\documentclass[runningheads]{llncs}

\input{Packages}

\title{Revisiting DRUP-based Interpolants with CaDiCaL 2.0}

\author{
  Basel Khouri\orcidlink{0009-0002-4875-4277}
  \and 
  Yakir Vizel\orcidlink{0000-0002-5655-1667}
}

\institute{%
  Technion, Computer Science Department, Haifa, Israel\\
  \email{basel.khouri@campus.technion.ac.il}\\
  \email{yvizel@cs.technion.ac.il}
}

\date{}
\begin{document}
\maketitle

\input{Sections/Abstract}
\input{Sections/Introduction}
\input{Sections/Preliminaries}
\input{Sections/Implementation}
\input{Sections/Results}
\input{Sections/FutureWork}

\subsubsection*{Acknowledgment} This research was partially supported by the Israeli Science Foundation (ISF) grant No. 2875/21.

\FloatBarrier

\bibliographystyle{splncs04}
\bibliography{references}

\end{document}

%% file: Packages.tex
\usepackage{macros}
\usepackage{paralist}
\usepackage{amssymb}
\usepackage{orcidlink}
\usepackage{siunitx}
\PassOptionsToPackage{hyphens}{url}\usepackage{hyperref}
\usepackage{amsmath}
\usepackage{cleveref}
\usepackage[utf8]{inputenc}
\usepackage[right]{lineno}
\usepackage{csquotes}
\usepackage{booktabs}
\usepackage{longtable}
\usepackage{adjustbox}
\usepackage{array}
\usepackage{url}
\usepackage{authblk}
\usepackage{float}
\usepackage{algorithm}
\usepackage{algpseudocode}
\usepackage{comment}
\usepackage{xspace}
\usepackage{listings}
\usepackage{xcolor}
\usepackage{subcaption}
\usepackage{multirow}
\usepackage{placeins}
\lstdefinestyle{cppstyle}{
  language=C++,                      % The language of the code
  basicstyle=\ttfamily\scriptsize,   % Reduced font size (\scriptsize or \tiny)
  keywordstyle=\color{blue},         % Color of C++ keywords
  stringstyle=\color{red},           % Color of strings
  commentstyle=\color{gray},         % Color of comments
  numberstyle=\tiny\color{gray},     % Style for line numbers
  stepnumber=0,                      % Line number step
  numbersep=8pt,                     % Distance of line numbers from code
  showspaces=false,                  % Show spaces in the code
  showstringspaces=false,            % Don't show spaces in strings
  tabsize=4,                         % Set tab size to 4 spaces
  breaklines=true,                   % Break long lines
  breakatwhitespace=false,           % Only break at whitespace
  captionpos=b,                      % Caption position at the bottom
  numbers=left,                      % Line numbers on the left
  frame=single,                      % Frame around the code
}

%% file: Sections/Abstract.tex
\begin{abstract}
We present our implementation of DRUP-based interpolants in \cadical 2.0, and evaluate performance in the bit-level model checker \avy using the Hardware Model Checking Competition benchmarks. 

\cadical is a state-of-the-art, open-source SAT solver known for its efficiency and flexibility.
In its latest release, version 2.0, \cadical introduces a new proof tracer API. This paper presents a tool that leverages this API to implement the DRUP-based algorithm for generating interpolants.

By integrating this algorithm into \cadical, we enable its use in model-checking workflows that require interpolants. Our experimental evaluation shows that integrating \cadical with DRUP-based interpolants in \avy results in better performance (both runtime and number of solved instances) when compared to \avy with \glucose as the main SAT solver.

Our implementation is publicly available and can be used by the formal methods community to further develop interpolation-based algorithms using the state-of-the-art SAT solver \cadical. Since our implementation uses the \tracer API, it should be maintainable and applicable to future releases of \cadical.

\end{abstract}

%% file: Sections/Introduction.tex
\section{Introduction}

Boolean \emph{satisfiability} (SAT) is the problem of deciding if a given logical formula has a satisfying assignment or not. Modern SAT solvers can produce either a satisfying assignment or a proof of unsatisfiability for satisfiable or unsatisfiable formulas, respectively. There has been a growing interest in certifying the results returned by SAT solvers~\cite{DBLP:conf/fmcad/HeuleHW13,DBLP:conf/sat/WetzlerHH14,DBLP:journals/corr/Heule16,DBLP:conf/cade/Cruz-FilipeHHKS17}. Unlike satisfying assignments, certifying unsatisfiable instances is more involving and requires modifications to the SAT solver. This has led to an extensive research suggesting different proof formats that are on the one hand easy to generate (namely, without modifying the SAT solver extensively and without negatively affecting its performance), and on the other hand are easy to check. 

In this paper we focus on proofs of unsatisfiability and their usage in SAT-based model checking (MC)~\cite{DBLP:journals/ac/BiereCCSZ03,DBLP:conf/cav/McMillan03,DBLP:conf/fmcad/VizelG09,DBLP:conf/vmcai/Bradley11,DBLP:conf/cav/VizelG14}. SAT-based MC reduces model checking to instances of SAT, where in most cases, a satisfiable instance indicates the existence of a counterexample, while an unsatisifiable instance corresponds to bounded correctness. Hence, in the case of an unsatisfiable instance, a proof of unsatisfiability corresponds to a bounded proof of correctness. These kind of proofs have many usages in model checking, such as in abstraction-refinement~\cite{DBLP:conf/tacas/McMillanA03}, interpolation~\cite{DBLP:conf/cav/McMillan03,DBLP:conf/fmcad/VizelG09,DBLP:conf/cav/VizelG14}, debugging, and more. 

We present an implementation of a framework that enables interpolant computation from DRUP proofs, which originally appeared in ~\cite{DBLP:conf/fmcad/GurfinkelV14}, using \href{https://github.com/arminbiere/cadical}{\cadical} as a SAT solver. \cadical is a state-of-the-art open-source SAT solver renowned for its efficiency and flexibility. Our implementation, \drupitp, leverages the new \tracer API introduced in \cadical 2.0~\cite{DBLP:conf/cav/BiereFFFFP24} to trace proof-related events during solver execution, enabling in-memory incremental DRUP proof logging.
Unlike existing proof tracers in \cadical, which are primarily designed for proof certification, \drupitp is tailored for model checking applications: it facilitates proof core extraction and interpolant computation, making \cadical 2.0 applicable to abstraction-based and SAT-based model checking algorithms.

\drupitp, presented in this paper, is more than a re-implementation of~\cite{DBLP:conf/fmcad/GurfinkelV14} in a modern SAT solver.
In addition, it incorporates various enhancements that are due to better engineering and more efficient algorithms.
From an engineering point of view, \drupitp is decoupled from the SAT solver and does not require any modifications to the SAT solver. This adds flexibility and allows the proof maintained by \drupitp to be different than the proof \cadical maintains, even during an incremental execution. This design is enabled mainly by the \tracer API. As a result, it allows us to develop a \emph{minimization} algorithm, which modifies the proof generated by the solver and ``prepares'' it for interpolation computation. This algorithmic improvement highly affects the performance of \drupitp.

The development of \drupitp enables the integration of \cadical as the underlying SAT solver in \avy~\cite{DBLP:conf/cav/VizelG14,DBLP:conf/cav/VizelGM15,DBLP:conf/cav/KrishnanVGG19}, a bit-level model checker that requires sequence interpolants. We evaluate how \drupitp with \cadical affects the performance of \avy using the HWMCC'19 and HWMCC'20 benchmarks. The results clearly demonstrate that the overall runtime performance of \avy, when using \cadical and \drupitp, is improved when compared to vanilla \avy (using \glucose with DRUP-based interpolants). Moreover, it can solve instances that the vanilla version cannot solve within a one-hour time limit. Furthermore, the proof minimization algorithm used in \drupitp improves interpolant generation when using \cadical incrementally. 

Lastly, our implementation of \drupitp and its integration in \avy, is publicly available\footnote{\url{https://github.com/TechnionFV/extavy_public}} and can be used by other members of the formal methods community to further improve interpolation techniques and their usage in model checking and other applications.

\subsection{Related Work}

Aside from the interpolation support described in~\cite{DBLP:conf/cav/BiereFFFFP24}, we are not aware of other tools that allow generating interpolants using \cadical. Unlike the approach we present in this paper, the generation of interpolants in~\cite{DBLP:conf/cav/BiereFFFFP24} is based on generating the antecedents by \cadical for every learned clause. This approach, conceptually, is similar to generating and using the resolution proof to compute the interpolant. Even if the entire resolution proof is not constructed and saved during the execution of the solver, and partial interpolants are computed on-the-fly per learned clause, this means that partial interpolants may be computed for clauses that do not participate in the derivation of the empty clause. Lastly, the antecedent-based approach may suffer from a similar performance issue in the case of incremental solving (as we noticed in our experiments, when the minimization technique is not enabled, see Section~\ref{sec:res:min}).

When compared to~\cite{DBLP:conf/fmcad/GurfinkelV14}, the implementation presented in this paper does not require any modifications to the internals of the SAT solver. This fact makes \drupitp easily maintainable and applicable to future releases of \cadical. Moreover, this separation enables application-specific modifications to the proof (e.g. minimization), without modifying the proof maintained by the SAT solver.

%% file: Sections/Preliminaries.tex
\section{Preliminaries}
\subsection{CNF, Resolution, and UP}
Given a set $U$ of Boolean variables, a \emph{literal} $\ell$ is either a
variable $u\in U$ or its negation $\neg u$, a \emph{clause} is a
disjunction of literals, and a formula in \emph{Conjunctive Normal Form} (CNF) is a conjunction of clauses.

The \textit{resolution rule} is a fundamental inference rule that captures many of the steps performed by SAT solvers. It allows deriving a new clause by resolving two clauses containing complementary literals.
Formally, applying the resolution rule on the clauses $C_1 = (A \lor l)$ and $C_2 = (B \lor \neg l)$, where $A$ and $B$ are clauses (that do not contain complementary literals) and $l$ and $\neg l$ are complementary literals, derives the clause $C_3 = (A \lor B)$. The derived clause $C_3$ is implied by $C_1 \land C_2$, thereby can be added to the CNF while preserving equivalence.

As defined in~\cite{DBLP:conf/fmcad/GurfinkelV14}, a \textit{resolution derivation} of a clause $\alpha$ from a CNF formula $G$ is sequence $\pi=(\alpha_1, \alpha_2,..., \alpha_n\equiv\alpha)$, where each clause $\alpha_k$ is either an original clause of $G$ or is derived by applying the resolution rule to clauses $\alpha_i, \alpha_j$ with $i,j<k$.
A resolution derivation $(\alpha_1, \alpha_2,..., \alpha_k)$ is \textit{trivial}, if all variables resolved on are distinct, and each $\alpha_i$, for $i > 2$, is either an original clause of $G$ or derived by resolving $\alpha_{i-1}$ with an original clause.
A \textit{chain resolution rule}, written $\alpha_1, \alpha_2, ..., \alpha_k \vdash^{\vec{x}}_{\text{TVR}}\alpha$ , states that $\alpha$ can be derived from $\alpha_1, \alpha_2, ..., \alpha_k$, the \textit{chain}, by a trivial resolution derivation, where $\vec{x}=(x_1,x_2,...,x_k)$ are the \textit{chain pivots}.
A \textit{chain derivation} is a sequence $\pi=(\alpha_1,\alpha_2,...,\alpha_n)$ where each $\alpha_k$ is either an original clause of $G$ or is derived by chain resolution from preceding clauses.
A
\emph{derivation witness} of a chain derivation $\pi$ is a total
function $D$ from clauses of $\pi$ to sub-sequences of $\pi$ such that
\begin{align}
  D(\alpha) = [\;] &\limp \alpha \text{ is initial} &
  D(\alpha) \neq [\;] &\limp D(\alpha) \vtres \alpha
\end{align}
A resolution proof of unsatisfiability of $G$ is a chain derivation of the empty clause from $G$.
We refer to a derivation of the empty clause as a \emph{proof}.

Given a partial assignment $\tau$, a clause $c$ is a unit clause under $\tau$ if there exists exactly one literal $l\in c$ such that $l,\neg l\not\in\tau$, while the rest of the literals in $c$ are assigned to false, i.e., $\forall l^{'}\in c, l^{'}\not = l: l^{'}\not\in\tau, \neg l^{'}\in\tau$. Given a formula $G$ and an assignment $\tau$,  \textit{Unit Propagation} (a.k.a \textit{Boolean Constraint Propagation}, or BCP) w.r.t. $G$ and $\tau$, is the process of repeatedly extending $\tau$ with unit literals from $G$ until a fixed-point is reached.
If for some literal $l$, both $l$ and $\neg l$ are in $\tau$, we say Unit Propagation \textit{derives a conflict}. 

\subsection{Validating Clausal Proofs}
\label{sec2.3}
In \cite{clausalproofs}, the authors show that the sequence of all learnt clauses, in the order they are learned, by a CDCL SAT solver, form a chain derivation. Moreover, they show that the chain derivation can be validated using Unit Propagation (UP) facilities of the solver. 

\begin{definition}[RUP]\label{def:rup}
    Let $G$ be a CNF formula and $c$ a clause over $\Var(G)$. If UP derives a conflict w.r.t. $G\land \neg c$ then $c$ is \emph{deducible from $G$}, denoted as  $G\vbcp c$. Clauses that are deducible via UP are also known to have the \emph{Reverse Unit Propagation} (RUP) property.
\end{definition}

The following lemma relates RUP and trivial resolution.

\begin{lemma}[\cite{21}]
Given a CNF formula $G$ and a clause $c$, $c$ is deducible from $G$ by UP iff $c$ is deducible from $G$ by trivial resolution. That is, $G\vbcp c \iff G\vtres c$.
\end{lemma}

Two algorithms were introduced in \cite{clausalproofs}: one for forward and one for backward proof validation. The forward validation algorithm replays the proof forward, checking that each clause is subsumed (using UP) by prior clauses. Backward validation walks the proof backwards, removing clauses, and checking that each removed clause has the RUP property w.r.t. the remaining clauses.

Backward proof validation was improved in~\cite{DBLP:conf/fmcad/HeuleHW13}, where a new proof clausal proof format, \textit{Delete Reverse Unit Propagation} (DRUP), was introduced. DRUP is a sequence \(\pi=((\alpha_0, d_0), ..., (\alpha_n, d_n)\) \(\equiv(\square, \bot))\), where each $d_k$ is a Boolean flag indicating whether the clause $\alpha_k$ is deleted, and $\alpha_k$ is either an original clause or is derived by chain resolution from the set of $k-$active clauses $\{\alpha_j|(j < k \land d_j = \bot) \land (\forall j < i < k: \alpha_i \not = \alpha_j)\}$.
Including clause deletion information makes validating a DRUP proof much more efficient since the validation of a clause $\alpha_k$ depends only on the $k-$active clauses.
In backward validation~\cite{DBLP:conf/fmcad/HeuleHW13}, it is possible to \textit{trim} the proof by removing all clauses that do not participate in the derivation of the empty clause.

\subsection{Interpolation and Colors}
\subsubsection{Craig Interpolant}
Given a pair of CNF formulae $\langle A,B\rangle$ such that $A \land B$ is unsatisfiable, \textit{Craig Interpolant} is a formula $I$ such that 
\begin{inparaenum}[(i)]
    \item $A \models I$,
    \item $I \models \neg B$, and
    \item $I$ is over the common variables of $A$ and $B$.  
\end{inparaenum}
It is known that an interpolant can be computed in polynomial time from a resolution proof of unsatisfiability of $\langle A,B\rangle$ \cite{mcmillan2003interpolation,a0092568-8f2a-3f18-91e1-717cd568fea3}.

\begin{definition}[$N$-Colored CNF]\label{def:colored_cnf}
    A \emph{$N$-colored CNF} is
a pair $(G, \kk)$ of a CNF formula $G$ and a coloring function $\kk :
G \to [1,\ldots,N]$ that assigns to every clause $\alpha \in G$ a color
between 1 and $N$. 
\end{definition}

The coloring extends naturally to
variables. For each $v \in \Var(G)$, we define its minimum and maximum
color as follows:
\begin{align}
  \kmin(v) &= \min \{ i \mid \exists \alpha \in G_i \cdot v \in \alpha
  \}\\
  \kmax(v) &= \max \{ i \mid \exists \alpha \in G_i \cdot v \in \alpha
  \}
\end{align}

\subsubsection{Sequence Interpolant}
A \emph{sequence interpolant} for an $N$-colored
unsatisfiable striped CNF $(G, \kk)$ is a sequence of formulas $(\top
\equiv I_0, \ldots, I_{N} \equiv \bot)$ such that for all $1 \leq i
\leq N$:
\begin{align*}
  I_{i-1} \land G_i &\limp I_{i} & 
  \forall v \in \Var(I_i) &\cdot \kmin(v) = i \land \kmax(v) = i+1
\end{align*}

\subsection{The Model Checker \avy}
\avy is a SAT-based model checker~\cite{DBLP:conf/cav/VizelG14,DBLP:conf/cav/VizelGM15,DBLP:conf/cav/KrishnanVGG19} that combines sequence interpolants~\cite{DBLP:conf/fmcad/VizelG09} and Property Directed Reachability (PDR)~\cite{DBLP:conf/vmcai/Bradley11,DBLP:conf/fmcad/EenMB11}. It uses bounded model checking (BMC)~\cite{DBLP:journals/ac/BiereCCSZ03} to search for a counterexample of length $k$. If such a counterexample does not exist, it extracts a sequence interpolant for the unsatisfiable BMC formula, and uses PDR to generalize the sequence interpolant and transform it to CNF. As a key part of the algorithm requires sequence interpolants, we use it in order to evaluate the DRUP interpolants framework implemented using \cadical.

%% file: Sections/Implementation.tex
\section{Implementing DRUP-based Interpolants in \cadical}

In this section we present the implementation of DRUP-based interpolation in \cadical.
In \cadical 2.0~\cite{DBLP:conf/cav/BiereFFFFP24}, a specialized mechanism is added for online interaction with proof-related operations performed by the solver. This is done via the \tracer API, which allows users to implement various techniques that relate to proof events (e.g. the addition or deletion of a clause), without modifying the SAT solver itself.
A user-defined tracer is connected to the internal solver of \cadical and can react online (via callbacks) to proof-related events. 

\subsection{The \drupitp class}

To implement the approach presented in~\cite{DBLP:conf/fmcad/GurfinkelV14}, we developed \drupitp, an extension of the \tracer\footnote{In Figure~\ref{fig:drup2itp}, the class extends \texttt{StatTracer}. \texttt{StatTracer} extends \tracer by adding statistics.} class. Figure~\ref{fig:drup2itp} presents a partial view, required for what follows, of the \drupitp class. \drupitp maintains a stack of DRUP clausal proofs and uses its own independent clause database, which is stored in a hash table (\clauses member). Additionally, it implements specialized Unit Propagation (UP) and Conflict Analysis procedures (\up and \analyze, respectively), and maintains its own \trail, \reasons, and watches data structures. These are required for an efficient implementation of \up.
Similar to the standard CDCL SAT algorithm~\cite{CDCL}, the trail is a queue that keeps track of all currently implied and assumed (decided) literals, in the order they were assigned. The watches data structure helps efficiently track two unassigned literals in each clause, allowing the solver to quickly identify when a clause becomes unit or falsified during UP. The \reasons data structure maps each assigned variable to the clause that caused its assignment (i.e., its reason clause).
It also features the \rup procedure, which performs Reverse Unit Propagation on a given clause and analyzes the resulting conflict (by calling \analyze). Since \drupitp logs a DRUP proof, every learned clause is guaranteed to have the RUP property, meaning that invoking \rup (correctly) always derives a conflict.

We note that there exists an implementation of the \tracer API which is used to check DRAT proofs. This implementation includes a hash table for storing clauses and implements unit propagation. \drupitp uses this implementation as a starting point but extends it to include interpolation specific features.

\begin{figure}
    \centering
\begin{lstlisting}[style=cppstyle, label={cpp:declare}]
class Drup2Itp : public CaDiCaL::StatTracer {
  ...
  vector<Clause *> proof;    // clausal proof
  vector<Clause *> reasons;  // reasons for literals on trail
  vector<int> trail;         // currently assigned literals
  Clause **clauses;          // hash table of clauses
  ...
  bool UP (bool core = false, unsigned color = 0);
  void RUP (Clause *, unsigned &);
  void analyze_core ();
  void set_current_partition (unsigned);
  bool trim (ItpClauseIterator *, bool incremental = true);
  bool replay (ResolutionProofIterator &, bool incremental = true);
};
\end{lstlisting}
    \caption{The \drupitp class}
    \label{fig:drup2itp}
    \vspace{-14pt}
\end{figure}

Figure~\ref{fig:clause_obj} presents the clause object used by \drupitp. It contains the clause’s literals, identifier (\id), \range, a pointer to the \texttt{next} clause in the hash table, and three Boolean flags: \original, \garbage, and \core. Upon each clause addition notification from the solver, the tracer API provides \drupitp with relevant details about the clause, including its \id, whether it is an original or learned clause, and its literals. Using this information, \drupitp creates a new clause object, assigns it the clause \id, literals, sets the \original flag, and stores it in its clause database. A clause can either be an \original clause from the SAT problem or a clause learnt by the solver. If \core flag is set to \true, the clause is part of the computed proof core, whereas if \garbage flag is set to \true, the clause is deleted (or "detached") --- it is no longer watched and is skipped during Unit Propagation (UP).

\begin{figure}
    \centering
\begin{lstlisting}[style=cppstyle, label={cpp:clause}]
class Range {
  unsigned m_min : 16, m_max : 16;
}

class Clause {
  Clause *next;
  uint64_t id;
  bool original, garbage, core;
  Range range;
  unsigned size;
  int literals[1];
};
\end{lstlisting}
    \caption{Clause object.}\label{fig:clause_obj}
    \vspace{-12pt}
\end{figure}    

As in the original algorithm presented in \cite{DBLP:conf/fmcad/GurfinkelV14}, in order to support sequence interpolants clauses of the input formula are assigned different colors, with each original clause assigned a specific color. The \texttt{Range} class corresponds to the pair $(\kmin, \kmax)$, hence, for original clauses, the range of a clause object (see Figure~\ref{fig:clause_obj}) is always a singleton. For example, if an original clause $c$ is assigned the color $k$, then its range is set to $(k,k)$ (since $\kmin(c)=\kmax(c)=k$). For learned clauses, the range is determined during the \replay procedure (as described in the next section) and reflects the range of colors of all the clauses used to derive that learned clause in the given derivation witness (following the definition of $\kmin$ and $\kmax$).
The user is responsible for providing the coloring information to \drupitp. It accomplishes this by setting a specific member in \drupitp via \texttt{set\_current\_color()} method to track the current color, ensuring that each original clause added has its range set to the current color (as a singleton range).
The pseudocode algorithm~\ref{alg:solve}, presented in section ~\ref{Demonstration}, illustrates how colors are used when interpolants are required via the use of \texttt{set\_current\_color()}.

Although \drupitp maintains an independent clause database, it remains synchronized with the solver's clause database via clause identifiers (\id). This guarantees that any clause object in \drupitp and a clause object in \cadical sharing the same \id always contain the same literals. As a result, \drupitp can efficiently look up the matching clause in its hash table by computing the hash index directly from that identifier.

The separation of the implementation of \drupitp from the solver itself provides several benefits, including non-intrusive integration with the solver, simplified development, and greater flexibility in manipulating the proof independently of the proof maintained by the solver in \cadical. Additionally, because \drupitp logs a DRUP clausal proof, it does not rely on \cadical to supply antecedent information.

\subsection{Generating Interpolants On-The-Fly}
The process of computing an interpolant involves two main steps: \begin{inparaenum}[(1)] \item \trim: the proof is trimmed, and a proof core is identified; \item \replay: the proof core is replayed, and interpolants are computed on-the-fly. 
\end{inparaenum}
The declarations of these methods appear in Figure~\ref{fig:drup2itp}.
We do not go into the details of the interpolation system used to produce interpolants. Instead, we focus solely on the implementation details. We emphasize that the user can choose an interpolation system~\cite{DBLP:journals/jsyml/Krajicek97,DBLP:conf/cav/McMillan03,DBLP:conf/vmcai/DSilvaKPW10,DBLP:conf/fmcad/GurfinkelV14} and implement it (through the \resproofit class, discussed later).

We now describe the operation of \drupitp during different stages of execution: initialization, solving, trimming and replaying. Throughout the following composition, we assume the SAT solver is given the CNF formula $\varphi$.

\subsubsection{Initialization}
During initialization, a \drupitp instance is created initializing only an empty hash table (\clauses) to store the clauses and an empty clausal proof stack (\dproof). It is then connected to a \cadical instance, which from here on, we refer to as the \textit{solver}. Initialization of \drupitp is performed right at the beginning before clauses of $\varphi$ are added to the solver.

\subsubsection{Solving}
While solving, \drupitp is notified by the solver of every new clause addition and every existing clause deletion.
Throughout the execution of the solver \drupitp tracks these notifications.
Namely, for a clause addition notification, it creates and stores a new clause object, while for a clause deletion notification it marks the corresponding clause as \garbage. All notifications are logged in the DRUP clausal proof stack.
In addition, any unit literals (from newly added unary clauses) are appended to the \trail, and their reason clauses are recorded in \reasons. These units are not propagated but only added to the \trail during the execution of the solver.

\subsubsection{Trimming The Proof}
In case the solver derives the empty clause, \drupitp is now required to trim the proof. Before \trim is executed, \drupitp propagates \trail and derives the conflict on its own. Note that \drupitp derives the same conflict the solver derives. This is because it mirrors the derivation performed by the solver, i.e. it propagates all units in the same order before it propagates any newly discovered units, if any.

Let us denote the DRUP proof generated by the solver (and stored in \drupitp.\dproof) as $\Pi=\langle c_1,\ldots,c_n\rangle$.
\trim is responsible for removing clauses that do not contribute the the derivation of the empty clause. This is achieved by marking clauses that participate in the derivation of the empty clause as core clauses.

To better understand how \trim works, recall that in a valid DRUP proof for a CNF formula $\varphi$, a learned clause $c_i$, where $1\leq i\leq n$ must satisfy the RUP property (Definition~\ref{def:rup}). Let us denote by $\alpha(\Pi, i)$ all clauses $c_1,\ldots,c_{i-1}$ that are not marked as \garbage. Then for every $i$ it holds that $\varphi, \alpha(\Pi, i)\vbcp c_i$. Equivalently, executing \up when the active clauses only include $\varphi$ and $\alpha(\Pi, i)$ must result in a conflict. This property is used while trimming the proof. 

The procedure begins by marking the empty clause as \core. Note that at this point, all original and learned clauses that are not marked as \garbage are active. \trim proceeds by traversing the DRUP proof stack \dproof backwards. Let us denote the clause that is currently being processed during backward traversal as $c$. If $c$ is marked with \garbage (i.e. it is deleted), it is activated (and thus revived). Otherwise, $c$ is deactivated. Deactivation of a clause requires updating the \trail. Hence, in case $c$ is a reason clause for a literal on the \trail, that literal, and all literals coming after it are removed from the \trail. Assume that $d$ is a reason clause for a literal $l$ removed from the trail during this process. It is important to note that if $d$ is marked as \core, then all other reason clauses for literals $l'\in d$ where $l'\neq l$ are marked as \core. 

If $c$ is marked as \core (and not \garbage) and not marked with \original, literals in $\neg c$ are added as assumptions to the \trail and \up is called. Once a conflict is reached (recall $c$ satisfies the RUP property), \drupitp performs \emph{conflict analysis} (all these steps are done by invoking \rup). During conflict analysis the chain resolution leading to the derivation of $c$ is identified and its clauses are marked as \core.

\subsubsection{Replaying The Proof}
\label{replay}
Once the proof is trimmed and the core is identified, \replay is responsible for traversing the DRUP proof core forward, applying local transformations on the proof, and constructing an interpolant on-the-fly. To facilitate this, \replay accepts an instance of \resproofit (see Figure~\ref{fig:drup2itp} and Figure~\ref{fig:proof_iter}), which it notifies with resolution steps.

\begin{figure}
    \centering
\begin{lstlisting}[style=cppstyle, label={lst:res}]
struct ResolutionProofIterator {
  ChainDerivation chain;
  virtual ~ResolutionProofIterator ();
  virtual void chain_resolution (int resolvent) = 0;
  virtual void chain_resolution (Clause *resolvent = 0) = 0;
};
\end{lstlisting}
    \caption{Proof iterator}
    \label{fig:proof_iter}
    \vspace{-14pt}
\end{figure}

In order to support interpolants, clauses in \drupitp are marked with colors (Definition~\ref{def:colored_cnf}). The color of clauses and variables determine the way in which interpolants are constructed and depend on the interpolation method of choice. This is beyond the scope of this paper and we refrain from discussing these details. Yet, different interpolation procedures can be implemented using different implementations of the abstract class \resproofit.

\replay traverses the DRUP proof forwards, skipping over clauses that are not marked as core. It notifies the \resproofit with every clause marked as \core. For a learned clause $c_i$, \replay invokes \rup and analyzes the conflict, similar to what is done in \trim. During the conflict analysis, the chain derivation for $c_i$ is traversed on-the-fly using \resproofit.

The class \resproofit (see Figure~\ref{fig:proof_iter}) offers two methods, both of which indicate a chain resolution step. To report the complete chain derivation, \replay uses these two methods to notify \resproofit of all chain resolutions in a specific order, ensuring that each new clause is only reported after all the clauses used to derive it have been reported. The method \texttt{chain\_resolution(Clause*)} is used to report a resolvent that is a learnt clause, whereas \texttt{chain\_resolution(int)} is used to report a resolvent that is a unit clause propagated on the trail. In both cases, the \texttt{chain} member of \resproofit contains all the clauses and pivots participating in the chain resolution that derives the given learnt clause. Note that \texttt{chain} is reset before every call to \texttt{chain\_resolution()}. The traversal concludes when \replay notifies \resproofit with a \texttt{nullptr} as a resolvent, indicating the empty clause. At this point, \resproofit traversed the complete chain derivation.

\subsection{Exploring The Space of Proofs}
The \tracer API provides an infrastructure that allows interacting with proof events generated by the solver, without modifying the solver. As a result, \drupitp implements procedures like \up and \analyze, even though such procedures exist in \cadical. In this context, it is important to note that a derivation witness is not unique, and different implementations of procedures such as \up and \analyze, can lead to different chain derivations of the empty clause.
For example, in model checking applications, smaller proofs are sometimes preferred. In this case, core clauses can be prioritized when executing \up during the trimming, as suggested in~\cite{DBLP:conf/fmcad/HeuleHW13}.

Another notable example, also in the context of model checking, focuses on the \emph{shape} of produced interpolants. Due to various reasons (beyond the scope of this paper), it is often desired for interpolants to be in CNF~\cite{DBLP:conf/hvc/ChocklerIM12,DBLP:conf/cav/VizelRN13,DBLP:conf/cav/VizelG14}. CNF interpolants can be derived from colorable chain refutations~\cite{DBLP:conf/fmcad/GurfinkelV14}. Since the SAT solver does not necessarily generate colorable chain refutations, one can try and \emph{reorder} the proof during \replay. In general, reordering the entire proof may be infeasible, hence, in our implementation of \replay we use a specialized propagation procedure called \textit{color-ordered propagation}. In color-ordered propagation, propagation is performed in iterations, where in the $i^{th}$ iteration, only $i$-colored clauses are propagated. This implementation of the propagation procedure is designed to increase the colorability of the constructed chain refutation, i.e., to heuristically and efficiently attempt to bring the proof closer to a colorable chain refutation, with the goal of producing interpolants in which part of the interpolant can be represented by a CNF. More details about this approach can be found in~\cite{DBLP:conf/fmcad/GurfinkelV14}.

These are only two examples, but many other heuristics can be experimented with due to this architecture.

\subsection{Incrementality and Proof Minimization}
\label{sec:minimizer}

In many applications, and in model checking in particular, the underlying SAT solver is used incrementally. Incrementally trimming and replaying the proof, however, presents several challenges. One issue is that if the DRUP proof is not compressed after each invocation of the solver, the cumulative DRUP proof can become quite large. This may negatively impact both memory usage and the time required for trimming and replaying. While the \tracer API provides flexibility, it makes compressing the proof and managing the internal clause database of the solver quite tricky. Moreover, implementing such functionality goes against the separation between the proof and the state of the internal solver.

For example, one might attempt to compress the clause database that is maintained by \drupitp after trimming, keeping only the clauses that are marked as \core. While this approach is theoretically sound, it is difficult in practice. Any modification to the \drupitp clause database must be mirrored in the internal solver's clause database. If the clause databases in \drupitp and the solver are not synchronized, the solver may learn new clauses using clauses that are not ``visible'' to \drupitp. Such a scenario makes trimming and replaying more complicated.

To address this issue, we introduce a new class called \minimizer. The \minimizer has two key members a solver object \minsolver and a \drupitp object \minitp. We now briefly describe its operation.

During \trim, the \minimizer is notified about all clauses that are marked as \core, and adds them to \minsolver. These notifications are handled by \clauseit that is passed to \trim (see Figure~\ref{fig:drup2itp}). When trimming is done, the \minimizer invokes \minsolver (to ``re-solve'' only the original core clauses), and then invokes \minitp.\trim and \minitp.\replay, to compute the interpolant.

It is worth noting that solving the core subset of the problem is often empirically much faster than solving the original problem.
This approach further minimizes the proof core, leads to a more compact interpolant, and avoids the major degradation involved in incremental \replay as demonstrated by our experimental evaluation (Section~\ref{sec:results}).

\subsubsection{UNSAT Core Extraction}
While somewhat unrelated to interpolants, the above can also be used to extract an UnSAT Core and minimize it, specifically when the solver is used incrementally. This is useful in many applications such \emph{proof-based abstraction}~\cite{DBLP:conf/tacas/McMillanA03} or \emph{counterexample-guided abstraction refinement}~\cite{DBLP:journals/jacm/ClarkeGJLV03}.

\subsection{Demonstration}
\label{Demonstration}
We present an example to demonstrate how \drupitp are used to compute an interpolant in Algorithm~\ref{alg:solve} and Algorithm~\ref{alg:get_itp}. Moreover, we show how \minimizer is used to compute interpolants in Algorithm~\ref{alg:get_min_itp}.
In the following pseudocode, we assume that the class \itpgen is a derived class from the base class \resproofit. It computes an interpolant iteratively through two pure virtual methods that represent resolution steps in the proof produced by \drupitp during \replay, with each method being triggered by a callback from \drupitp.

\input{Pseudocode/Algorithm}

Recall that \replay traverses a derivation witness for the empty clause, and notifies the \resproofit instance of each learned clause and its chain resolution, ensuring that a clause is only reported after all the clauses used to derive it have been reported. This order is particularly well-suited for SAT-based interpolation algorithms that compute interpolants from a resolution proof such as McMillan interpolation system~\cite{mcmillan2003interpolation}.

As explained in Section~\ref{sec:minimizer}, a freshly allocated \minimizer is passed to the main \drupitp object, which then trims the instance and notifies the \minimizer of the original core. The \minimizer then proceeds to solve, trim, and replay only the original core. 

It is important to note that, in practice, minimization can involve more than one step. This means adding additional trim and solve steps, and other heuristics, to further minimize the proof that is be replayed for interpolation. However, for simplicity, we include only one minimization step in the pseudo-code.

%% file: Pseudocode/Algorithm.tex
\begin{algorithm}[t]
\caption{Set a \drupitp instance to trace a \cadical instance to solve a colored-CNF formula and compute an interpolant.}\label{alg:solve}
\begin{algorithmic}[1] % The [1] sets line numbering
    \State \textbf{Input:} An ordered set of CNF formulae $F$ representing partitions of an input CNF formula.
    % \State \textbf{Output:} Interpolant $I$ for the pair $\langle A,B \rangle$.
    \Function{\texttt{ITP}}{\texttt{vector<CNF> F}}
        \State \texttt{solver = new \cadical}
        \State \texttt{tracer = new \drupitp}
        \State \texttt{solver.connect\_proof\_tracer(\&tracer)}
        \State \texttt{i = 1}
        \For {\texttt{$P \in F$}}
            \State \texttt{tracer.set\_current\_color(i)}
            \For {\texttt{$c \in P$}}
                \State \texttt{solver.clause(c)}
            \EndFor
            \State \texttt{i = i+1}
        \EndFor
        \If {\texttt{solver.solve() == UNSAT}}
            \State \Return \texttt{getInterpolant(tracer)}
        \EndIf
    \EndFunction
\end{algorithmic}
\end{algorithm}

\begin{algorithm}[t]
\caption{Interpolant computation with \drupitp.}\label{alg:get_itp}
\begin{algorithmic}[1] % The [1] sets line numbering
    \State \textbf{Input:} A \drupitp object.
    % \State \textbf{Output:} Interpolant $I$ for the pair $\langle A,B \rangle$.
    \Function{getInterpolant}{\texttt{tracer}}
        \State \texttt{tracer.trim()}
        \State \texttt{interpolator = new \itpgen}
        % \State \textcolor{blue}{// Replay the trimmed proof and compute $I$ on-the-fly}
        \State \texttt{tracer.replay(interpolator)}
        \State \Return \texttt{interpolator.get\_interpolant()}
    \EndFunction
\end{algorithmic}
\end{algorithm}

\begin{algorithm}[t]
\caption{Interpolant computation with \minimizer.}\label{alg:get_min_itp}
\begin{algorithmic}[1] % The [1] sets line numbering
    \State \textbf{Input:} A \drupitp object.
    % \State \textbf{Output:} Interpolant $I$ for the pair $\langle A,B \rangle$.
    \Function{getInterpolant}{\texttt{tracer}}
        \State \texttt{minimizer = new \minimizer}
        \State \texttt{tracer.trim(\&minimizer)}
        \State \texttt{result = minimizer.solver.solve()}
        \State \texttt{assert(result = UNSATISFIABLE)}
        \State \texttt{minimizer.tracer.trim()}
        \State \texttt{interpolator = new \itpgen}
        % \State \textcolor{blue}{// Replay the trimmed proof and compute $I$ on-the-fly}
        \State \texttt{minimizer.tracer.replay(interpolator)}
        \State \Return \texttt{interpolator.get\_interpolant()}
    \EndFunction
\end{algorithmic}
\end{algorithm}

%% file: Sections/Results.tex
\section{Experimental Evaluation}
\label{sec:results}

\drupitp is implemented in our fork of \cadical\footnote{\url{https://github.com/TechnionFV/cadical_itp}}. By default, \cadical emits a DRAT proof. However, in the latest version available at the time of writing (2.1.2), \cadical does not actually produce RAT clauses, so we can safely treat the emitted proof as a DRUP proof. If future releases of \cadical introduce techniques that add RAT clauses to the proof, those features should either be disabled to ensure a a DRUP proof is emitted, or specialized support for DRAT proofs should be implemented.

\subsection{Avy}\label{sec:res:avy}
In order to evaluate \drupitp, we integrated \cadical 2.0~\cite{DBLP:conf/cav/BiereFFFFP24} with \drupitp into \avy~\cite{DBLP:conf/cav/VizelG14,DBLP:conf/cav/VizelGM15,DBLP:conf/cav/KrishnanVGG19}, a bit-level model checker that combines sequence interpolants and Property Directed Reachability (PDR)\footnote{\url{https://github.com/TechnionFV/extavy_public}}. Both \drupitp and its integration in \avy can be found on our group's \texttt{GitHub} page\footnote{\url{https://github.com/TechnionFV}}. Prior to the integration of \cadical and \drupitp in \avy, it supported only two SAT solvers --- \texttt{Minisat}~\cite{DBLP:conf/sat/EenS03} and \glucose~\cite{DBLP:conf/ijcai/AudemardS09}, both using DRUP-based interpolants~\cite{DBLP:conf/fmcad/GurfinkelV14}. The introduction of \drupitp allowed \avy to utilize \cadical 2.0, a state-of-the-art SAT solver.

We conducted a series of experiments comparing the performance of \avy across various configurations, using benchmarks from the Hardware Model Checking Competitions (HWMCC'19 includes 317 instances and HWMCC'20 includes 324 instances). The vanilla configuration of \avy uses \glucose (including its DRUP-based interpolants) as a SAT solver. All the other configurations uses \cadical as a SAT solver. In all \cadical-based configurations, when solving the BMC formula, \cadical is invoked with pre-processing and in-processing techniques. We tested several different configurations that affect \drupitp. Namely, with respect to sequence-interpolant computation:
\begin{enumerate}
    \item \drupitp with and without minimization.
    \item \minimizer with and without \emph{pre/in-processing}. 
\end{enumerate}

The same \avy switches are used for all different SAT solver configurations.
All experiments were executed on machines with AMD EPYC 74F3 CPU and 32GB of memory. The timeout was set to 3600 seconds.

\begin{table}[h!]
\scalebox{0.9}{
    \centering
    \begin{tabular}{c|c|c|c|c|c|c}
    \hline\hline
        Set        & Solver           & SAT Instances & UNSAT Instances  & Unique & Avg Runtime & Avg Depth \\
        \hline
        \multirow{3}{*}{HWMCC'19} 
                   & \cadical         & \bf 31      & 172           & 2     & 1504.6      & \bf 21.2  \\
                   & \minimizer       & 27          & \bf 179       & \bf 4 & \bf 1427.6  & 22.4   \\
                   & \texttt{Glucose} & 29          & 176           & 2     & 1447.5      & 25.5   \\
                   \hline
                   & Virtual Best     & \it 32      & \it 184       & \it 8 & \it 1337.4  & \it 19.2    \\
        \hline
        \hline
        \multirow{3}{*}{HWMCC'20} 
                   & \cadical         & \bf 34     & 169           & 4 & 1501.1      & \bf 19.2 \\
                   & \minimizer       & 31         & \bf 175       & \bf 6 & \bf 1446.5  &  19.8   \\
                   & \texttt{Glucose} & 33         & 168           & \bf 6 & 1511.6      & 23.0   \\
                   \hline
                   & Virtual Best     & \it 36     & \it 182       & \it 16 & \it 1363.8  & \it 18.0       \\
                   \hline
                   \hline
        
    \end{tabular}
    }
    \caption{\centering Evaluation results. Average runtime is in seconds, and Average Depth represents the average convergence depth for \avy.}
    \label{tab:benchmark}
    \vspace{-18pt}
\end{table}

Table~\ref{tab:benchmark} summarizes the results. The row labeled with \cadical is for \cadical operating in incremental mode with pre/in-processing enabled. The row labeled \minimizer is the same as \cadical with the addition of proof minimization (see Section~\ref{sec:minimizer}). In the presented configuration \minimizer is used with pre/in-processing techniques enabled.
This configuration yields the best results for \avy with respect to the benchmark set we used.
The results demonstrate that compared to the base version, \avy with \cadical and minimization performs better overall and solves instances that other configurations are unable to solve within a one-hour time limit. 

Another notable observation was that different configurations of \cadical with \drupitp solved different sets of instances, leading to a higher virtual best score when combined. This suggests that there is significant potential in tailoring configurations for specific types of instances.

\begin{figure}[t]
    \centering
    \begin{subfigure}{0.47\textwidth}
        \centering
        \includegraphics[width=1.0\textwidth]{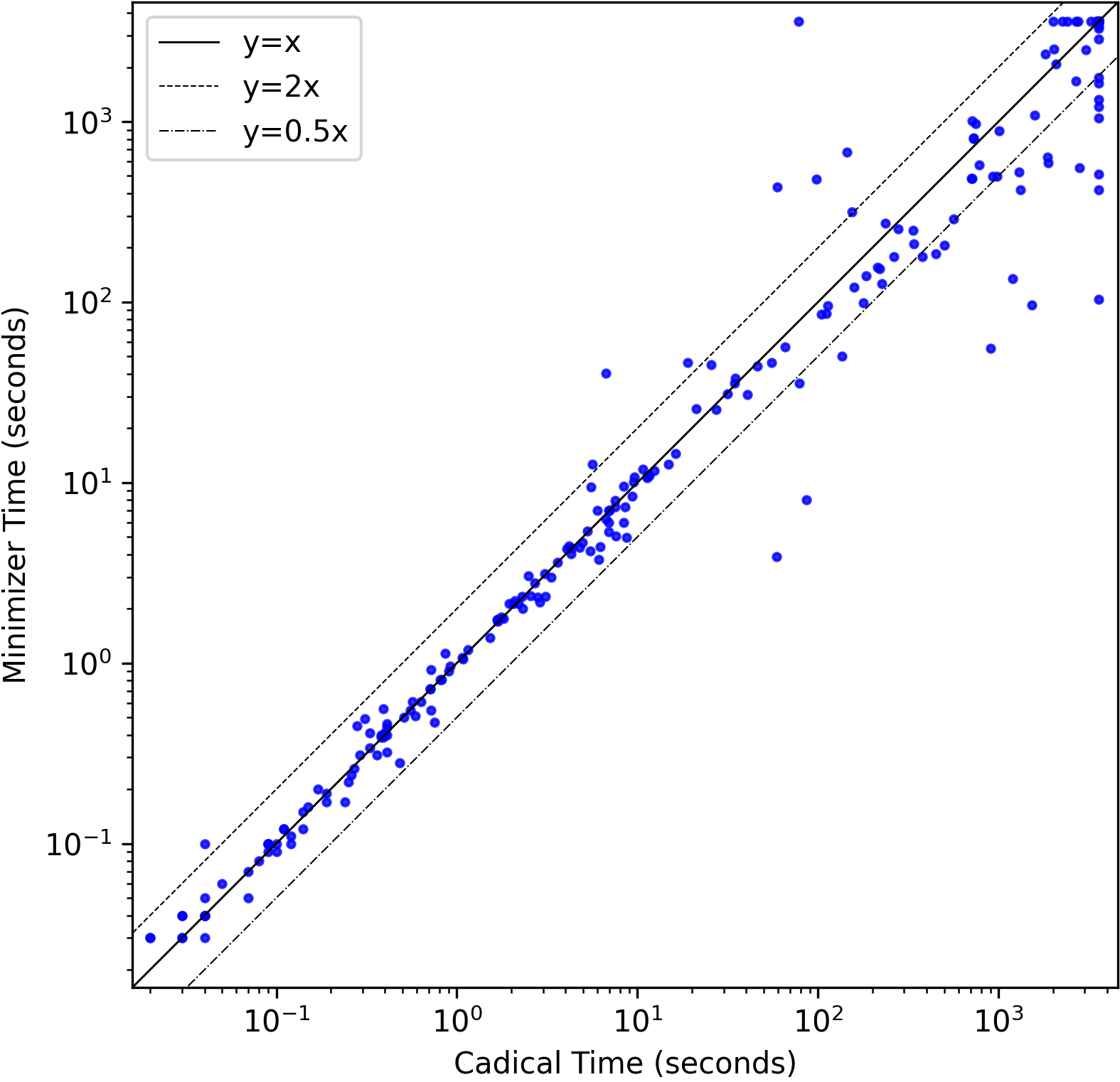}
        \caption{HWMCC'19}
        \label{fig:hwmcc19_minimize}
    \end{subfigure}
    \hfill
    \begin{subfigure}{0.47\textwidth}
        \centering
        \includegraphics[width=1.0\textwidth]{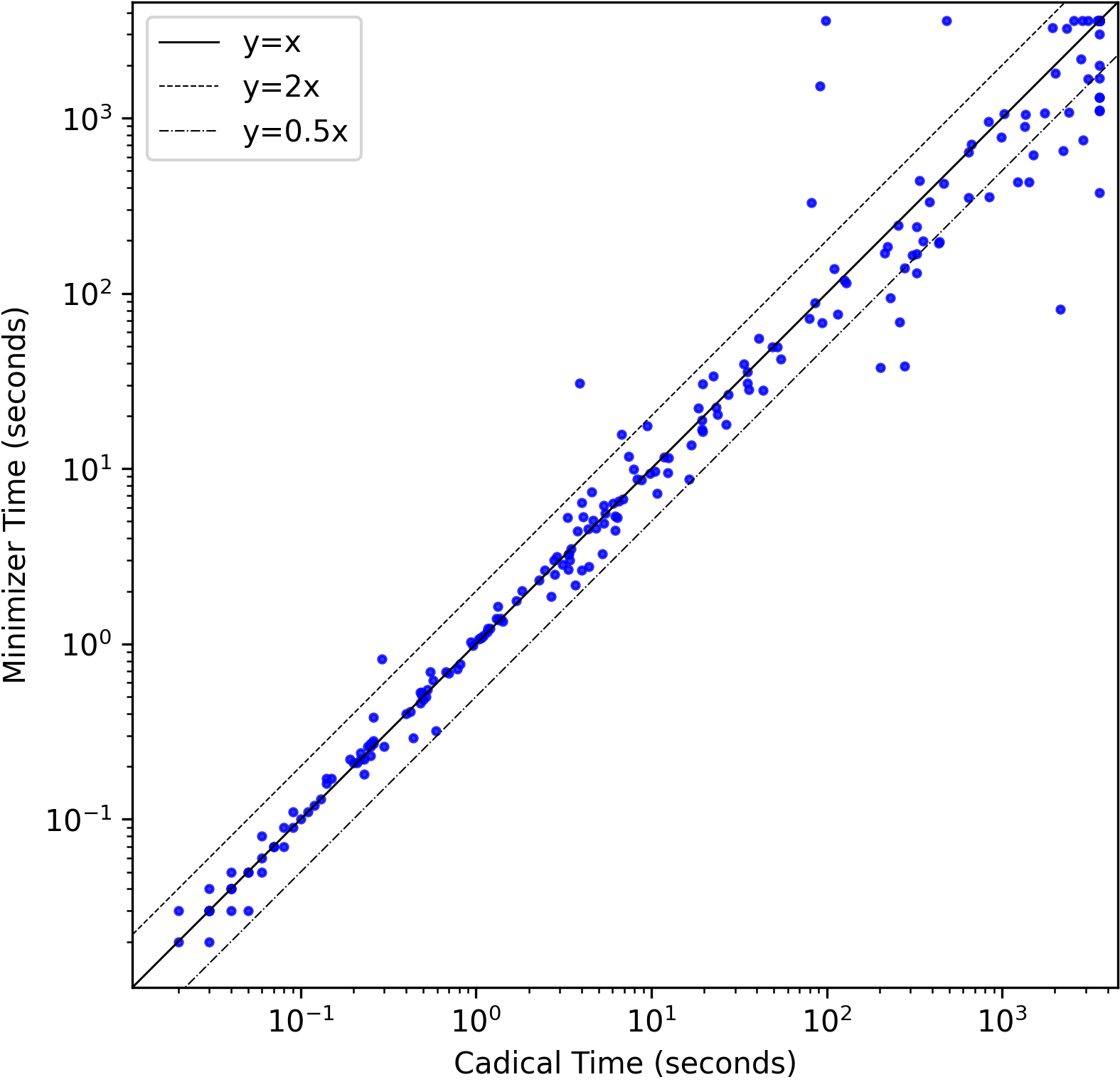}
        \caption{HWMCC'20}
        \label{fig:hwmcc20_minimize}
    \end{subfigure}
    \caption{Comparing performance with and without \minimizer.}
    \vspace{-14pt}
\end{figure}

To further illustrate the impact minimization has on performance, we compare the runtime of \cadical and \minimizer. This comparison appears in the scatter plots in Figure~\ref{fig:hwmcc19_minimize} and Figure~\ref{fig:hwmcc20_minimize}, for both HWMCC'19 and HWMCC'20, respectively. The results indicate that using \minimizer leads to better runtime performance as the majority of instances are placed under the diagonal of parity.

\subsection{Incrementality and Minimization}\label{sec:res:min}
In Section~\ref{sec:minimizer}, we discuss the degradation in interpolant computation time and size associated with incremental solving in \drupitp, and explain how the \minimizer method addresses it. To illustrate this, we use HWMCC'19 and HWMCC'20 benchmarks, and run the two variants described above: \cadical and \minimizer. During this run, we force the \avy model checker to compute sequence interpolants at every bound. Specifically, for each bound $k$, both variants unroll the model to depth $k$, solve a $BMC(k)$ instance and compute a sequence interpolant. Consequently, at every bound $k$, both variants start the process of computing a sequence interpolant from the same DRUP proof. The only difference is that the \minimizer invokes \replay on the minimized proof.

\begin{figure}[ht]
    \centering
    \begin{subfigure}{0.45\textwidth}
        \centering
        \includegraphics[width=1.0\textwidth]{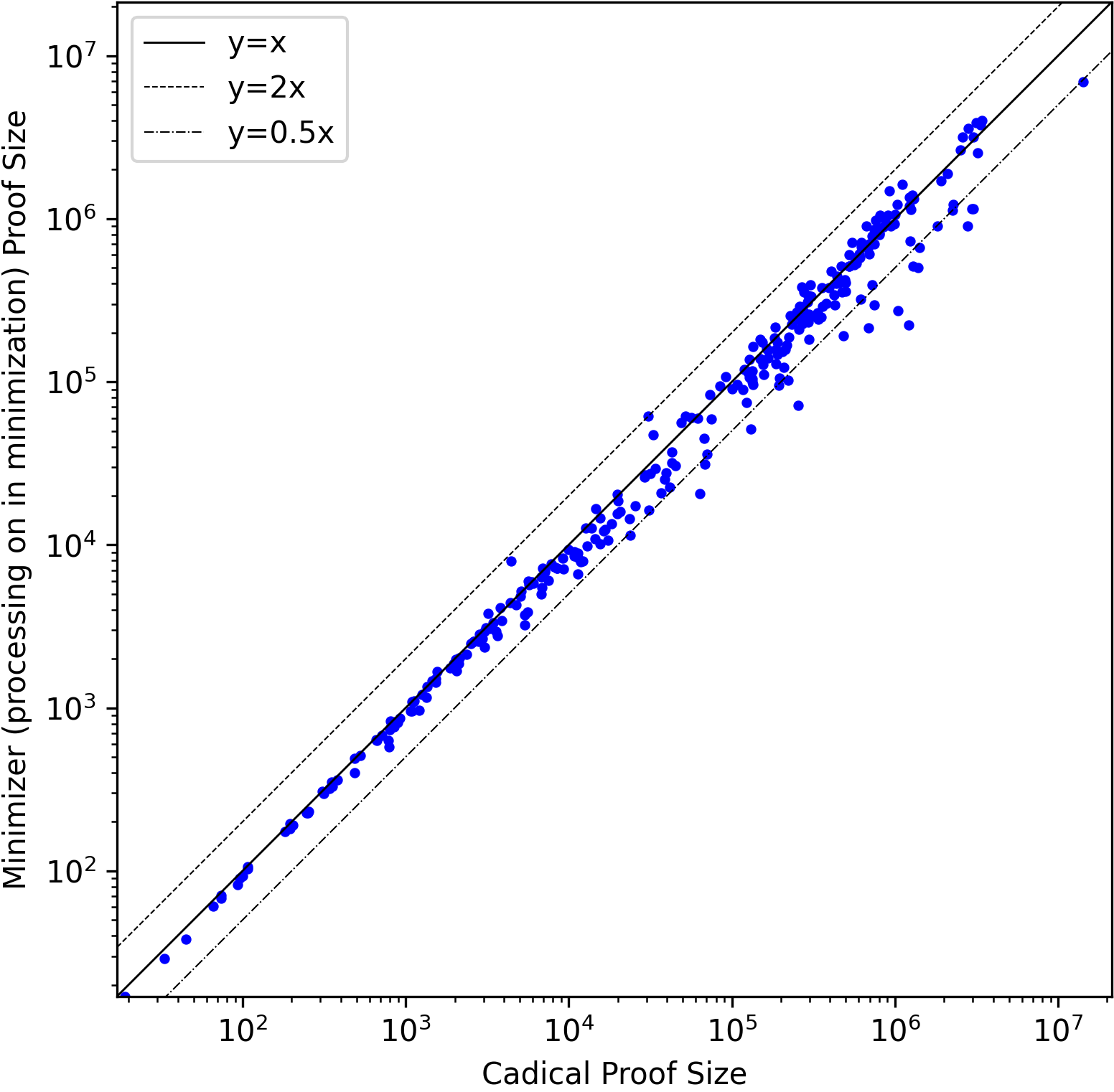}
        \caption{HMWCC'19 proof size comparison}
        \label{fig:19_proof_size}
    \end{subfigure}
    \hfill
    \begin{subfigure}{0.45\textwidth}
        \centering
        \includegraphics[width=1.0\textwidth]{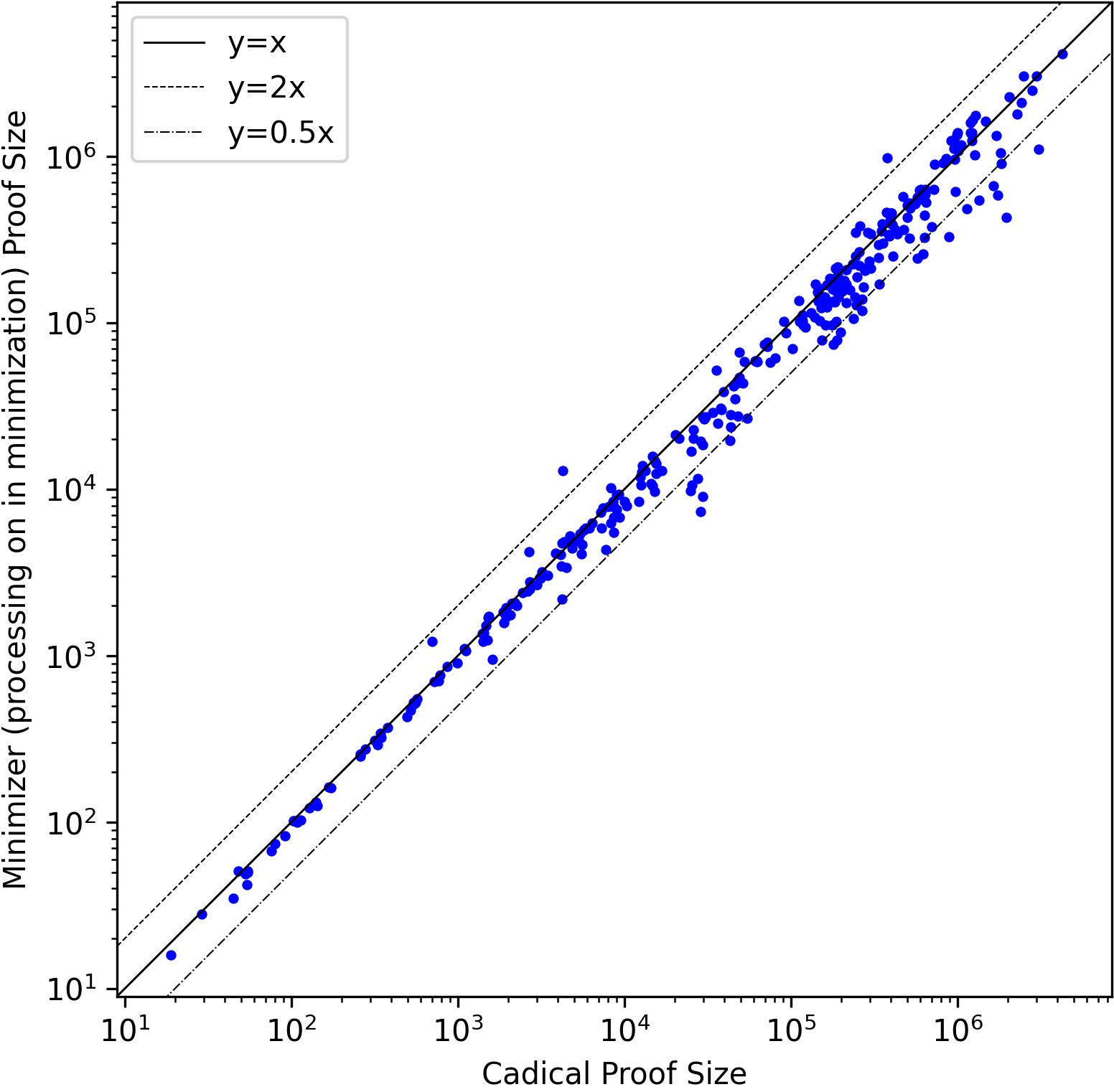}
        \caption{HMWCC'20 proof size comparison}
        \label{fig:20_proof_size}
    \end{subfigure}
    \hfill
    \begin{subfigure}{0.45\textwidth}
        \centering
        \includegraphics[width=1.0\textwidth]{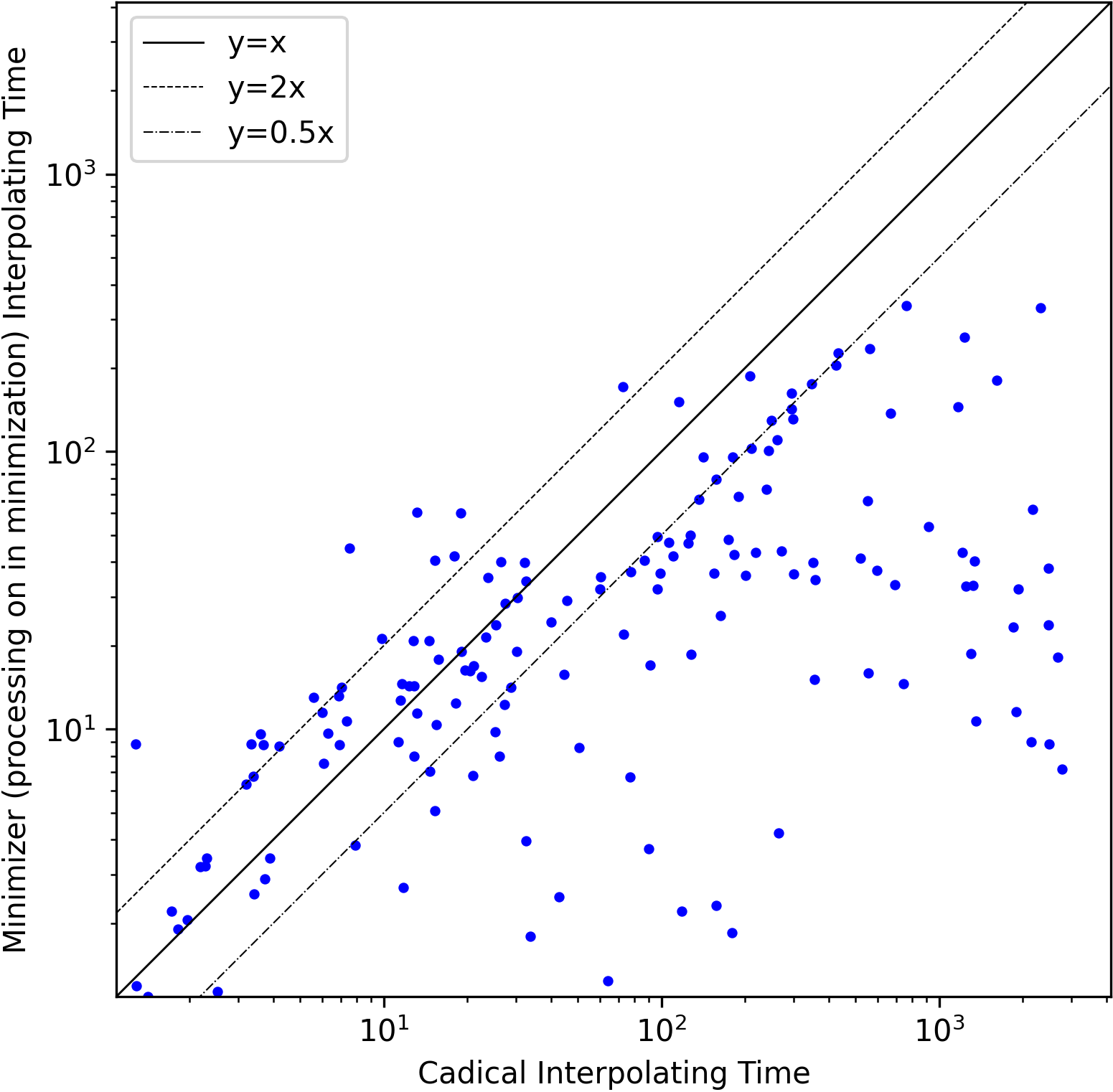}
        \caption{HMWCC'19 interpolant computation time comparison}
        \label{fig:19_itp_time}
    \end{subfigure}
    \hfill
    \begin{subfigure}{0.45\textwidth}
        \centering
        \includegraphics[width=1.0\textwidth]{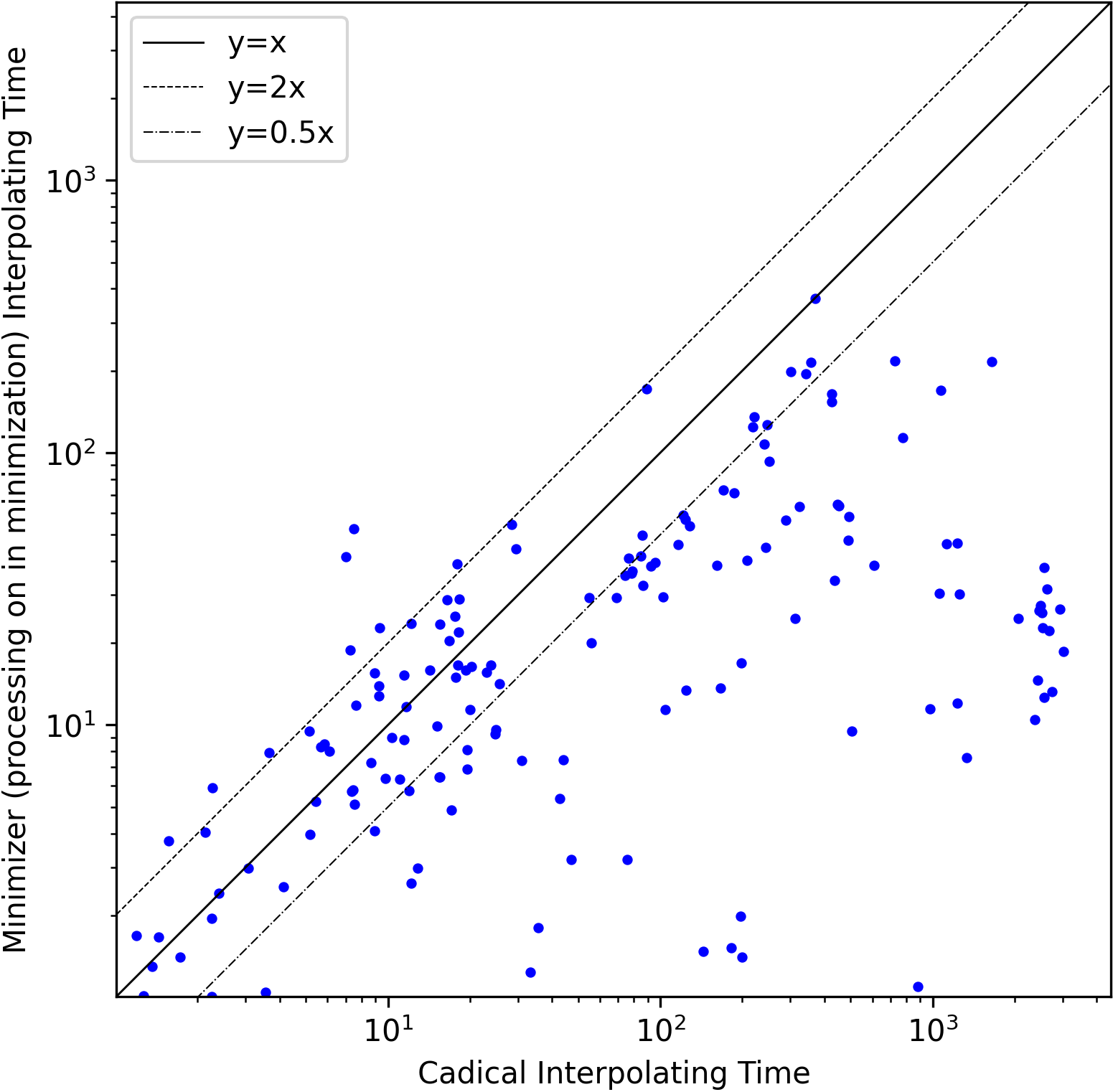}
        \caption{HMWCC'20 interpolant computation time comparison}
        \label{fig:20_itp_time}
    \end{subfigure}
    \hfill
    \begin{subfigure}{0.45\textwidth}
        \centering
        \includegraphics[width=1.0\textwidth]{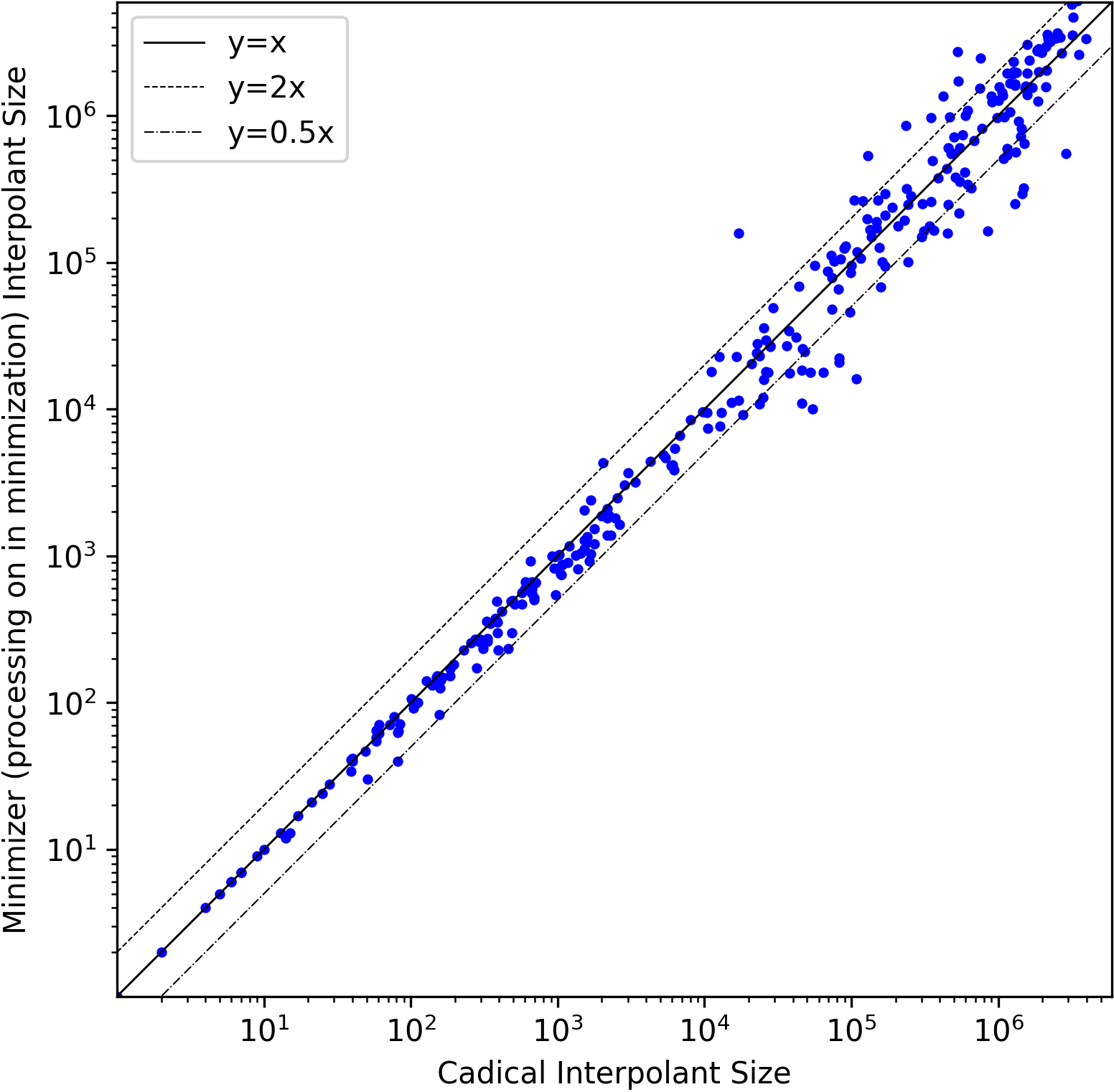}
        \caption{HMWCC'19 interpolant size comparison}
        \label{fig:19_itp_size}
    \end{subfigure}
    \hfill 
    \begin{subfigure}{0.45\textwidth}
        \centering
        \includegraphics[width=1.0\textwidth]{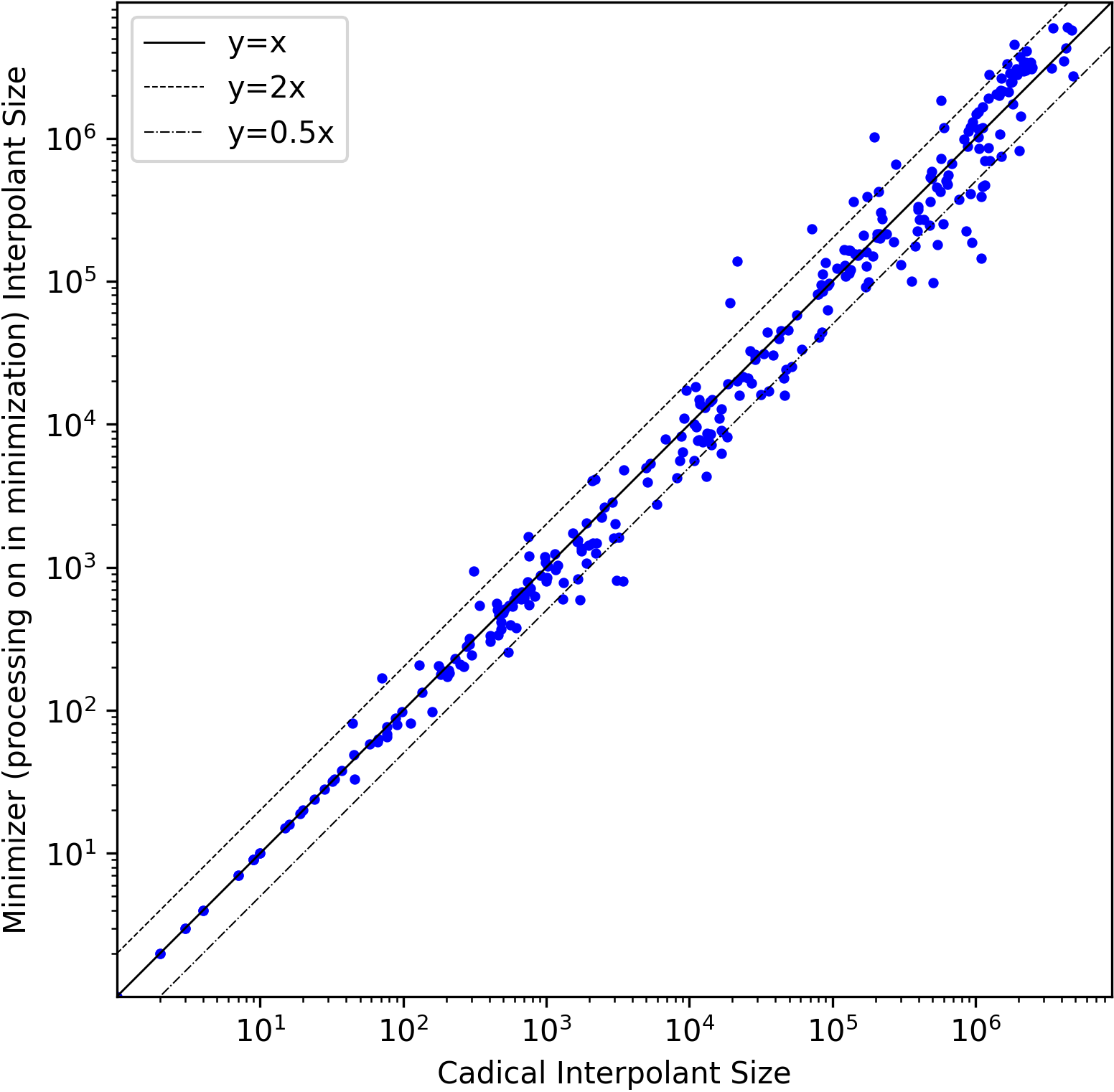}
        \caption{HMWCC'20 interpolant size comparison}
        \label{fig:20_itp_size}
    \end{subfigure}
    \caption{Proof size, interpolant size, and interpolant computation time comparison.}
    \vspace{-5pt}
\end{figure}

For each test and each bound $k$, we record both the replayed proof size and the resulting interpolant size (after simplification). In Figures \ref{fig:19_proof_size} and \ref{fig:20_proof_size}, we show, for each test, the sum of proof size among all bounds for both variants. Similarly, we report the sum of interpolant size in Figures \ref{fig:19_itp_size} and \ref{fig:20_itp_size}, and the sum of time it takes to compute an interpolant in Figures \ref{fig:19_itp_time} and \ref{fig:20_itp_time}. To keep comparisons fair, we compute values over mutual bounds and ignore any additional bounds one variant completes beyond the other within the same one-hour time limit. For instance, if the \minimizer variant executes BMC up to bound 200 but the \cadical variant only reaches 150, we average over the first 150 bounds only.

\begin{figure}[ht]
    \centering
    \begin{subfigure}{0.45\textwidth}
        \centering
        \includegraphics[width=1.0\textwidth]{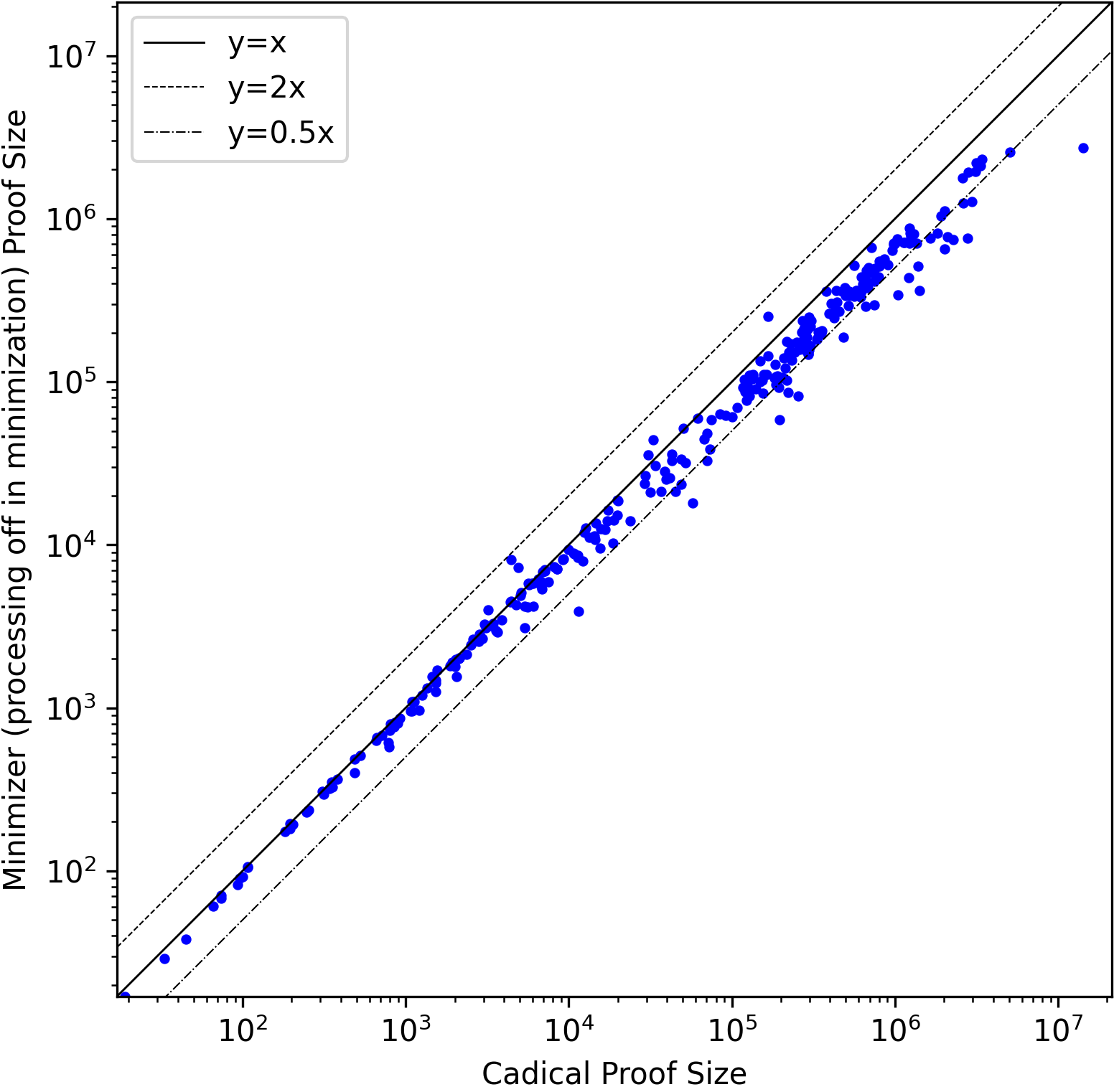}
        \caption{HMWCC'19 proof size comparison}
        \label{fig:19_proof_size_min_proc_off}
    \end{subfigure}
    \hfill
    \begin{subfigure}{0.45\textwidth}
        \centering
        \includegraphics[width=1.0\textwidth]{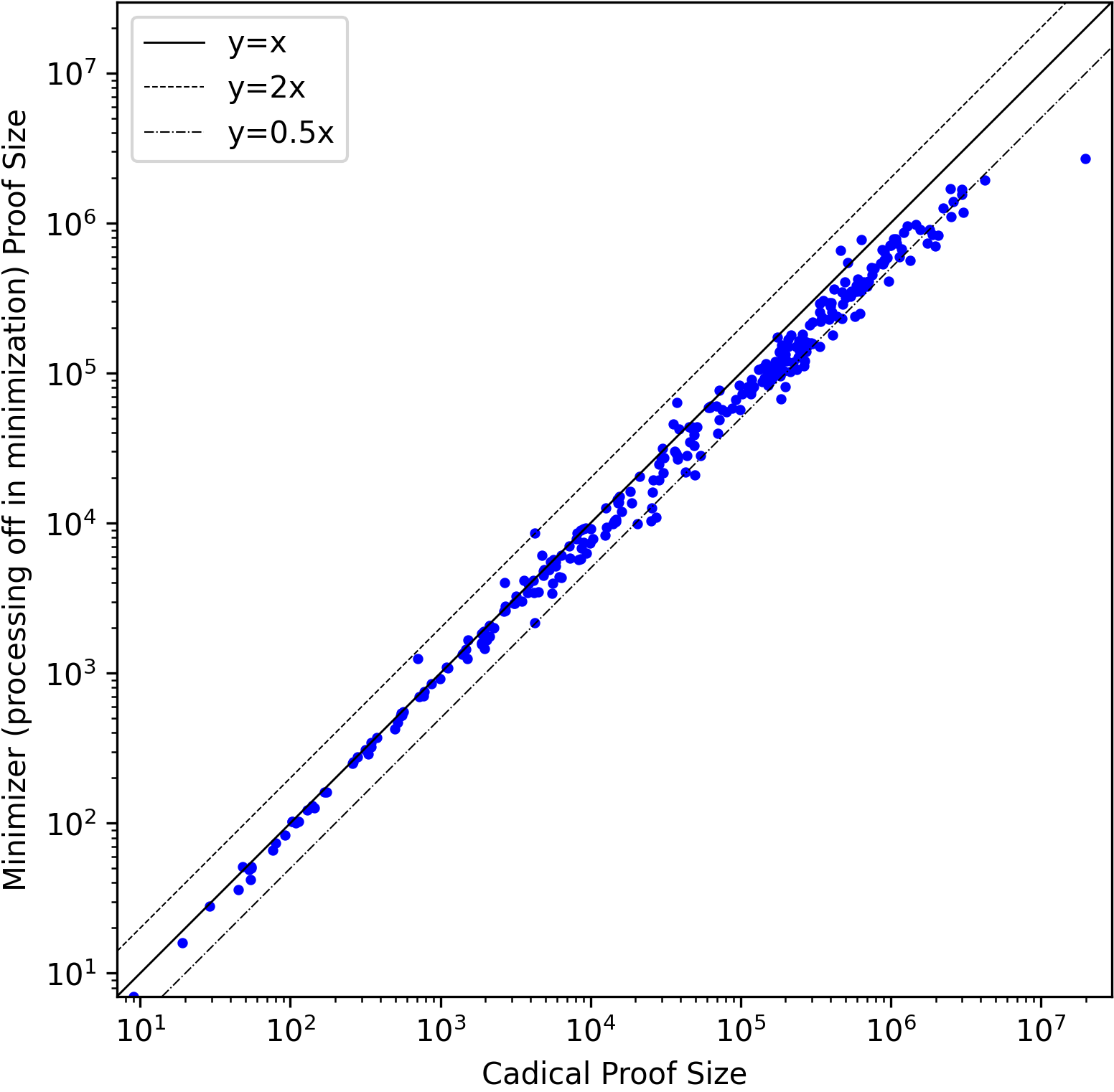}
        \caption{HMWCC'20 proof size comparison}
        \label{fig:20_proof_size_min_proc_off}
    \end{subfigure}
    \caption{Proof size comparison while disabling pre-/in-processing during minimization.}
    \vspace{-14pt}
\end{figure}

Analyzing the results, we observe that the replayed proof from which an interpolant is computed is significantly smaller in the \minimizer variant than in the \cadical variant.
Moreover, in $70\%$ of the cases, the resulting interpolant is, on average, smaller in \minimizer than in \cadical.
Furthermore, Figures \ref{fig:19_itp_time} and \ref{fig:20_itp_time} show that overhead of the \minimizer is not only minimal, but also highly beneficial, reducing the total runtime on average. While \cadical simply replays the main trimmed proof, \minimizer re-solves a core subset of the original clauses, trims the resulting proof, and then replays this minimized proof. These results also suggest that \replay time strongly correlates with the size of the replayed proof.

Another observation is that, by disabling pre-/in-processing during the minimization step (i.e., in the second solve call of each interpolant computation), \minimizer produces even smaller proofs, as shown in Figures \ref{fig:19_proof_size_min_proc_off} and \ref{fig:20_proof_size_min_proc_off}. However, although this approach remains competitive, its overall performance is slightly behind the \minimizer variant that keeps pre-/in-processing enabled when run on the \avy benchmarks from Section~\ref{sec:res:avy}.
This suggests that enabling pre-/in-processing in the minimization step creates a trade-off between proof size and solve time. Nevertheless, the resulting interpolants did not consistently show a comparable reduction in size.

%% file: Sections/FutureWork.tex
\section{Conclusions}

In this work we present an implementation of DRUP-based interpolants in \cadical 2.0. Our implementation uses the \tracer API and therefore does not require any modifications to the internals of the SAT solver. This architecture allows us (and others) to further explore possible improvements to interpolants computation in \cadical.

As future work, we plan to further develop this framework and extend it such that it supports other proof formats (e.d. DRAT and LRAT). Moreover, we intend to further investigate how model checking algorithms, that require interpolants, can further be improved, harnessing the full potential of modern SAT solvers such as \cadical.